\documentclass[reprint, superscriptaddress, amsmath,amssymb, aip, apl, floatfix, 10pt]{revtex4-1}

\usepackage[utf8]{inputenc}
\usepackage[english]{babel}
\usepackage{microtype}
\usepackage{graphicx}
\usepackage{hyperref}
\usepackage{color}

\newcommand{\blu}{\color{black}}   

\begin{document}

\title{Manipulating photon coherence to enhance the security of practical \\quantum key distribution}

\author{George L. Roberts}
\email{glr28@cam.ac.uk}
\affiliation{Toshiba Research Europe Ltd, 208 Cambridge Science Park, Milton Road, \\Cambridge CB4 0GZ, United Kingdom}%
\affiliation{Cambridge University Engineering Department, 9 J J Thomson Avenue, \\Cambridge, CB3 0FA, United Kingdom}%
\author{Marco Lucamarini}%
\affiliation{Toshiba Research Europe Ltd, 208 Cambridge Science Park, Milton Road, \\Cambridge CB4 0GZ, United Kingdom}%
\author{James F. Dynes}%
\affiliation{Toshiba Research Europe Ltd, 208 Cambridge Science Park, Milton Road, \\Cambridge CB4 0GZ, United Kingdom}%
\author{Seb J. Savory}%
\affiliation{Cambridge University Engineering Department, 9 J J Thomson Avenue, \\Cambridge, CB3 0FA, United Kingdom}
\author{\\Zhiliang Yuan}%
\affiliation{Toshiba Research Europe Ltd, 208 Cambridge Science Park, Milton Road, \\Cambridge CB4 0GZ, United Kingdom}%
 \author{Andrew J. Shields}%
\affiliation{Toshiba Research Europe Ltd, 208 Cambridge Science Park, Milton Road, \\Cambridge CB4 0GZ, United Kingdom}%

\date{\today}

\begin{abstract}
Quantum key distribution (QKD) allows two users to communicate with theoretically provable secrecy by encoding information on photonic qubits. 
Current encoders are complex, however, which reduces their appeal for practical use and introduces potential vulnerabilities to quantum attacks. 
Distributed-phase-reference (DPR) systems were introduced as a simpler alternative, but have not yet been proven practically secure against all classes of attack. 
Here we demonstrate the first DPR QKD system with information-theoretic security. 
Using a novel light source, where the coherence between pulses can be controlled on a pulse-by-pulse basis, we implement a secure DPR system based on the differential quadrature phase shift protocol. 
The system is modulator-free, does not require active stabilization or a complex receiver, and also offers megabit per second key rates, almost three times higher than the standard Bennett-Brassard 1984 (BB84) protocol.
This enhanced performance and security highlights the potential for DPR protocols to be adopted for real-world applications.
\end{abstract}

\maketitle

Quantum key distribution (QKD) has developed strongly since the proposal of the first protocol in 1984~\cite{Bennett_quantum_1984, gisin_quantum_2002, scarani_security_2009}.
The future could see widespread quantum networks similar to those in Tokyo~\cite{sasaki_field_2011} and Vienna~\cite{peev_secoqc_2009} and global secure communication enabled by QKD over satellites~\cite{vallone_experimental_2015}.
These advances depend on the development of simple, cost-effective and high performance implementations.
Innovations in both protocols and system hardware are required to achieve this.

Nearly two decades after the inception of Bennett-Brassard 1984 (BB84)~\cite{Bennett_quantum_1984}, distributed phase reference (DPR) QKD was proposed, allowing for much simpler experimental implementations.
The class includes the differential phase shift~\cite{Inoue_differential_2002, Takesue_quantum_2007} and coherent-one-way~\cite{Stucki_fast_2005, Korzh_provably_2015} protocols.
One advantage is that the transmitters needed to realize these DPR protocols can be made using off-the-shelf telecom lasers and modulators.
However the benefit of their simpler implementation is outweighed by a seriously degraded performance when full security is taken into account~\cite{scarani_security_2009, Moroder_security_2012, Mizutani_Information-theoretic_2017}.
To plug the security gap, two further DPR protocols were proposed: round-robin differential phase shift and differential quadrature phase shift (DQPS).
The former simplifies the estimation of Eve's information, but requires an overly complicated QKD receiver setup~\cite{Wang_experimental_2015, Takesue_experimental_2015, Guan_experimental_2015, Liu_round-robin_2017}, making it impractical.
The latter separates the signal from the differential phase shift protocol into blocks, each having a global phase that varies randomly, ensuring the protocol is immune against coherent attacks~\cite{Inoue_differential-quadrature-phase-shift_2009, kawakami_security_2016}.
It does, however, stray from the main goal of DPR protocols to provide simpler QKD implementations, due to the phase randomization requirement that would ordinarily require extra system components.

In this work we show it is possible to produce phase coherent and phase randomized pulses from a single device.
This device is based on optical injection of one laser diode into another, removing the need for a phase-randomization component in DQPS by relying on the randomness provided by spontaneous emission~\cite{yuan_directly_2016}.
This transmitter is stable, has a small footprint and allows us to achieve a base quantum bit error rate (QBER) of just 2.15\%.
We obtain a secure key rate of 2.37 Mbit/s at short distances, and show positive key rates up to an equivalent distance of 110 km.
The secure keys rates measured using real optical fiber channels align well with those obtained using an optical attenuator.
We also compare the secure key rates obtained with both protocols and find that, on average, DQPS produces keys at a rate 2.71 higher than BB84.

The differential phase shift protocol was the first DPR protocol proposed.
In this system, Alice encodes one of two random orthogonal phase values onto a coherent stream of pulses.
Bob then measures the bit values using an interferometer, inferring the presence of Eve by a break in coherence of the pulses during communication.

The DQPS protocol splits the differential phase shift signal into blocks of length L.
Each of these blocks has a globally random phase, which removes the coherence between pulses in different blocks.
Four phases are used in two non-orthogonal bases.
These act as the data Z \{0,~$\pi$\} and check X \{$\pi$/2,~3$\pi$/2\} bases.
We note that with a block size L=2, the DQPS protocol is identical to the phase-encoded BB84 protocol.

For implementation, the protocol starts with Alice randomly deciding her encoding basis for each block and bit value for each pulse inside the block.
She gives each block a globally random phase before sending them to Bob.
Bob uses an MZI with a one-bit time delay to measure the phase of each pulse in a randomly determined basis for each block.
If Bob detects a photon in a block, he discards any other photon clicks that occur at a later time in the same block.
If both detectors click at the same time, he randomly assigns a measurement.
Bob announces when he measured a pulse in each block, allowing Alice to determine a raw key.
They then announce which basis they used for each pulse, allowing them to share a sifted key and then perform error correction and privacy amplification.

A security proof is outlined by Kawakami \textit{et al}~\cite{kawakami_security_2016} that draws on a modified tagging technique and the complementarity argument.
Using this, the extracted asymptotic key rate is given by
\begin{equation}
\label{eq:keyRate}
R = \frac{n_{\textrm{rep}} p_0^2 Q}{L} \left[ 1 - f_{\textrm{PA}}(Q , E_1) - f_{EC}\left( \frac{E_0}{Q} \right) \right],
\end{equation}
where the privacy amplification factor is
\begin{equation}
\label{eq:privAmp}
f_{\textrm{PA}}(Q, E_1) = \frac{r_{\textrm{tag}}}{Q} + \left(1 - \frac{r_{\textrm{tag}}}{Q}\right) \textrm{h} \left( \frac{E_1}{Q-r_{\textrm{tag}}} \right),
\end{equation}
the probability that Alice emits more than one photon in adjacent pulses is
\begin{equation}
\label{eq:rTag}
r_{\textrm{tag}}=1-\sum_{m=0}^{L/2} e^{-\mu L}\mu^m \frac{(L+1-m)!}{m!(L+1-2m)!},
\end{equation}
$n_{\textrm{rep}}$ is the repetition rate of the source laser, $p_0$ is the probability of Alice preparing a state in the data basis, Q is the total gain, L is the block length and $E_{0,1}$ are the errors in the data and check basis respectively.
h(x) is the binary entropy function truncated to 1 at x values over 0.5 and the error correction factor $f_{EC}(E_0/Q)=\textrm{h}(E_0/Q)$.

Due to its small block size, the BB84 protocol can implement phase randomization in a straightforward manner.
A gain-switched pulsed laser can ensure perfect phase randomization~\cite{Yuan_robust_2014}, while an asymmetric Mach-Zehnder interferometer (MZI) provides the necessary block size~\cite{Dixon_high_2015}.
The increased block sizes required by the DQPS protocol effectively prevents the interferometer-based solution because stabilizing a large number of interferometer arms is a formidable task.
An alternative approach would be to use a phase modulator for active block-wise phase randomization~\cite{Zhao_experimental_2007}, which is attractive in theory but problematic in practice.
It would require a high-speed source of perfectly random numbers and infinitely precise electrical modulation signals.
We note that the DQPS protocol has not yet been demonstrated, despite its conceptual simplicity.

\begin{figure}[b]
    \centering
    \includegraphics[width=0.9\linewidth]{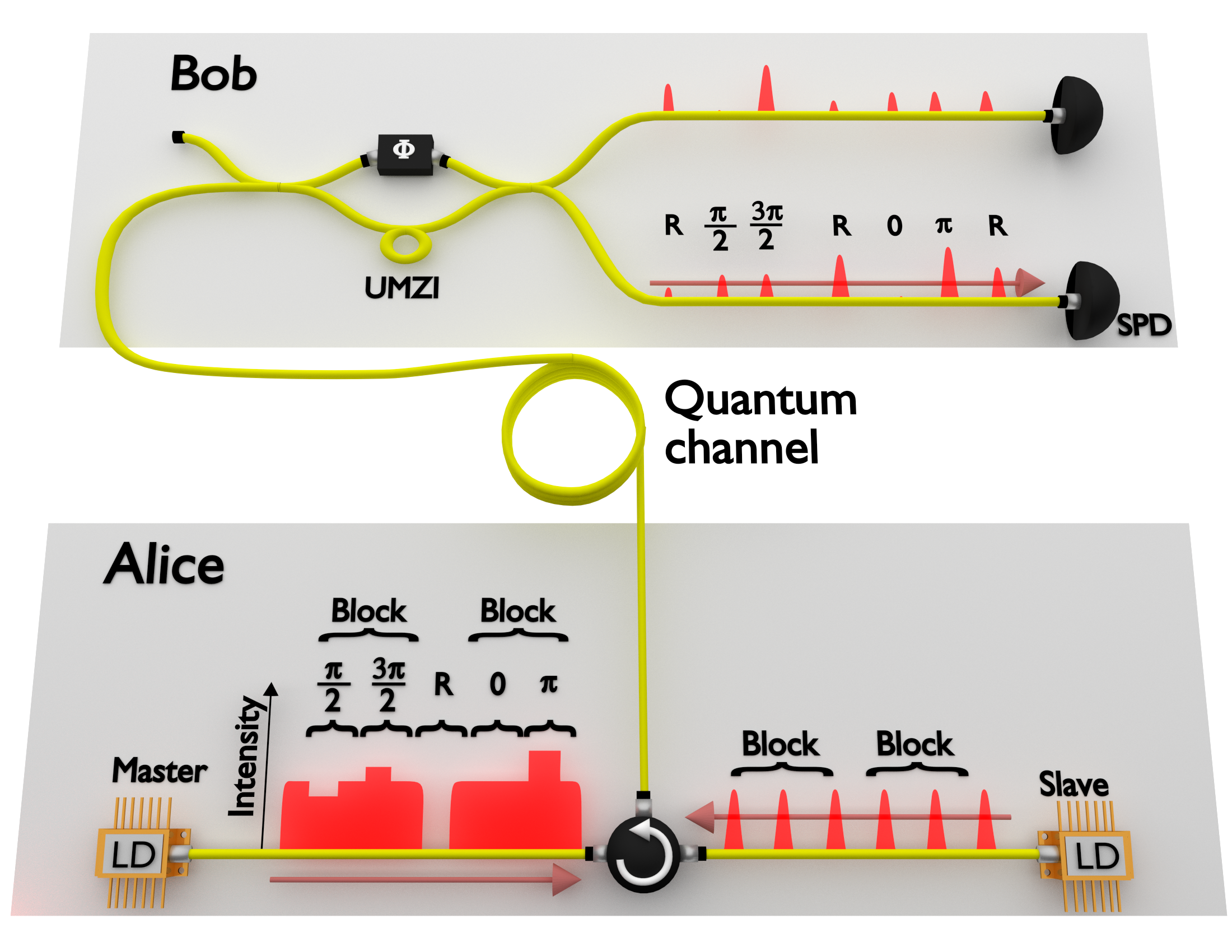}
    \caption{\textbf{Experimental setup for the DQPS protocol.} A master laser diode (LD) injects phase modulated light into a 2~GHz gain switched slave laser diode via a circulator.
    We draw L=3 here, however an arbitrary block size can be set by applying the correct driving signal to the master laser.
    This is sent to Bob, who interferes the received pulses using an interferometer with a one-bit time-delay and a measurement basis selectable using a phase modulator, $\Phi$, {\blu from $\{0,\pi/2\}$, on one arm.
    Our implementation is proof-of-principle, so uses a thermal phase shifter in one arm. }
    The values he will detect are overlaid on the pulses, with R corresponding to a pulse with a random phase.}
    \label{fig:Setup}
\end{figure}

We implement the DQPS protocol with the directly-modulated light source~\cite{yuan_directly_2016} shown in Fig.~\ref{fig:Setup}.
A slave laser emits a gain-switched train of pulses, whilst a master laser controls the phase of the pulses.
A small modulation in the master laser applied temporally between adjacent slave laser pulses allows the phase of the pulses to be precisely controlled without affecting their frequency or intensity.
This design produces a transmitter that is compact compared to other phase modulation systems, and also features a low power consumption and high stability.
This transmitter has previously been demonstrated with established QKD protocols~\cite{yuan_directly_2016, Roberts_modulator-free_2017}.

Alice generates a 512-bit pseudo-random pattern and then assigns a basis to each block based on the probability of sending a `check' and `data' block.
Knowing the half-wave voltage of the system, modulations are applied in-between pulses in order to encode the desired phase shifts.
This output is passed through a polarization controller to align the light with Bob's MZI, then through a DWDM to filter out unwanted noise.
She attenuates her signal to the desired mean photon number.
The optimal mean photon number is calculated using a simulation based on Equation~\ref{eq:keyRate} for each experimental distance, which is also used to optimize the block size.
Larger block sizes and a lower mean photon number give better secure key rates at longer distances.
The block size is constrained to containing 2$^n$ useful pulses in order to match the pattern size, so L=2$^n$+1.
She sends the signals to Bob through the quantum channel, which is simulated by an attenuator for some measurements, and using standard optical fiber with a loss of 0.2~dB/km for others.

Bob uses a planar lightwave circuit MZI with a 500~ps time delay on one arm and a heater to select the measurement basis.
This component has an inherent 3~dB loss.
In our experiment, we use a superconducting nanowire single photon detector (SPD) with a total efficiency of 38.6~\% and a dark count rate of 15~Hz.
The low dark count rate ensures we are not limited by noise at long distances.
The experiment is proof-of-principle, so we measure data in each basis separately, until at least 400,000 counts are detected in both bases.
In a real experimental implementation, the basis could be chosen actively for each block, by using a high-speed phase modulator in one arm of the MZI.
The output of the SPD is interpreted by a digitizer with 100~ps time bins and a constant fraction discriminator to minimize detection time-jitter.
The detectors, laser diodes and MZI are independently temperature controlled, but no active feedback is given to the system during data collection.

\begin{figure}[t]
    \centering
    \includegraphics[width=\linewidth]{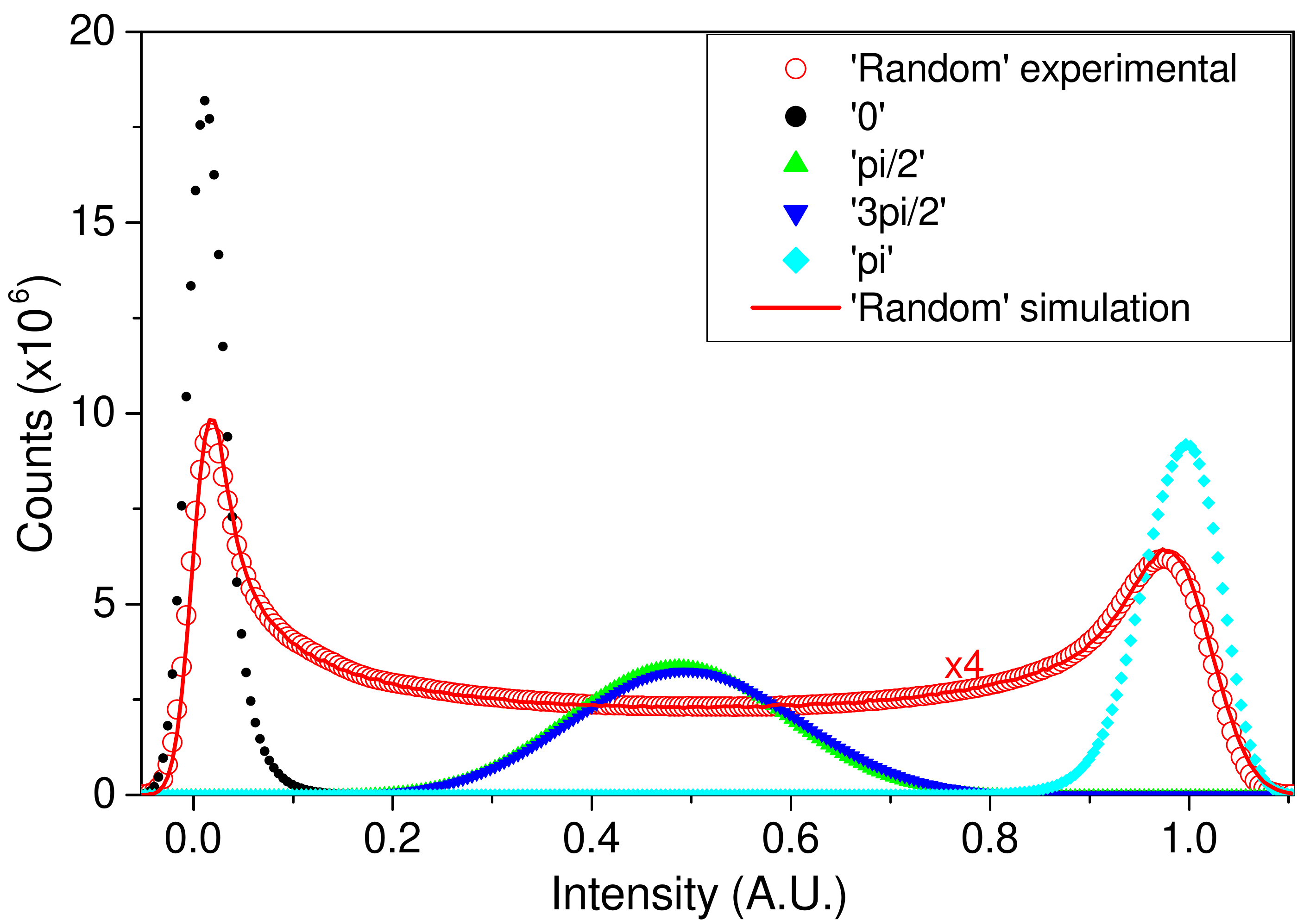}
    \caption{\textbf{Randomness of blocks.} Histogram of measured intensities for all DQPS signal values after the MZI. Experimental (symbols) data is given and a simulation (line) shows the expected inter-block interference results. The simulation accounts for experimental uncertainties and intensity fluctuations. All of the potential modulation values are plotted and the random signal spans the whole range. The MZI is aligned to measure in the Z basis. 1.95$\times$10$^8$ samples are taken for each signal value and the random counts are multiplied by four for visibility.}
    \label{fig:randomness}
\end{figure}

The transmitter in Fig.~\ref{fig:Setup} enables global phase randomization of arbitrarily large block sizes with ease.
It does not need an extra phase modulator and random number generator.
 The phase continuity of the master laser can be disrupted by driving it below its threshold for a short period of time.
A duration of 125~ps is sufficient to deplete the laser cavity field, forcing the subsequent laser pulse to inherit a completely random phase from spontaneous emission.
In this regime, the evolution within each block is continuous, but is completely random between master emission blocks.
Therefore, we are able to achieve both intra-block phase modulation and inter-block phase randomization.
After inputting a DQPS pattern, we measure the output intensities from a one-bit interferometer, where all four modulation values are shown alongside the random interference between blocks in Fig.~\ref{fig:randomness}.
A simulation of the expected inter-block interference intensity shows excellent agreement with the experimental data.
We perform an autocorrelation measurement on inter-block interference data and observe that the results are distributed evenly within the expected confidence bounds, further confirming the block randomness.
This autocorrelation measurement is shown in Fig.~S1 of the supplementary material.

The probability of Bob detecting a `click' in a given time-slot is given by $P_{\textrm{click}}^1$=$n/n_{\textrm{rep}}$, where n is the number of valid detections.
From this we can calculate Q, defined as the probability of having just one click in a block:
\begin{equation}
Q=1-\left(1-P_{\textrm{click}}^1\right)^{L-1}.
\end{equation}
We use this value alongside the measured QBER and Equations~\ref{eq:keyRate}--\ref{eq:rTag} to calculate our secure key rates.

\begin{figure}[ht]
    \centering
    \includegraphics[width=\linewidth]{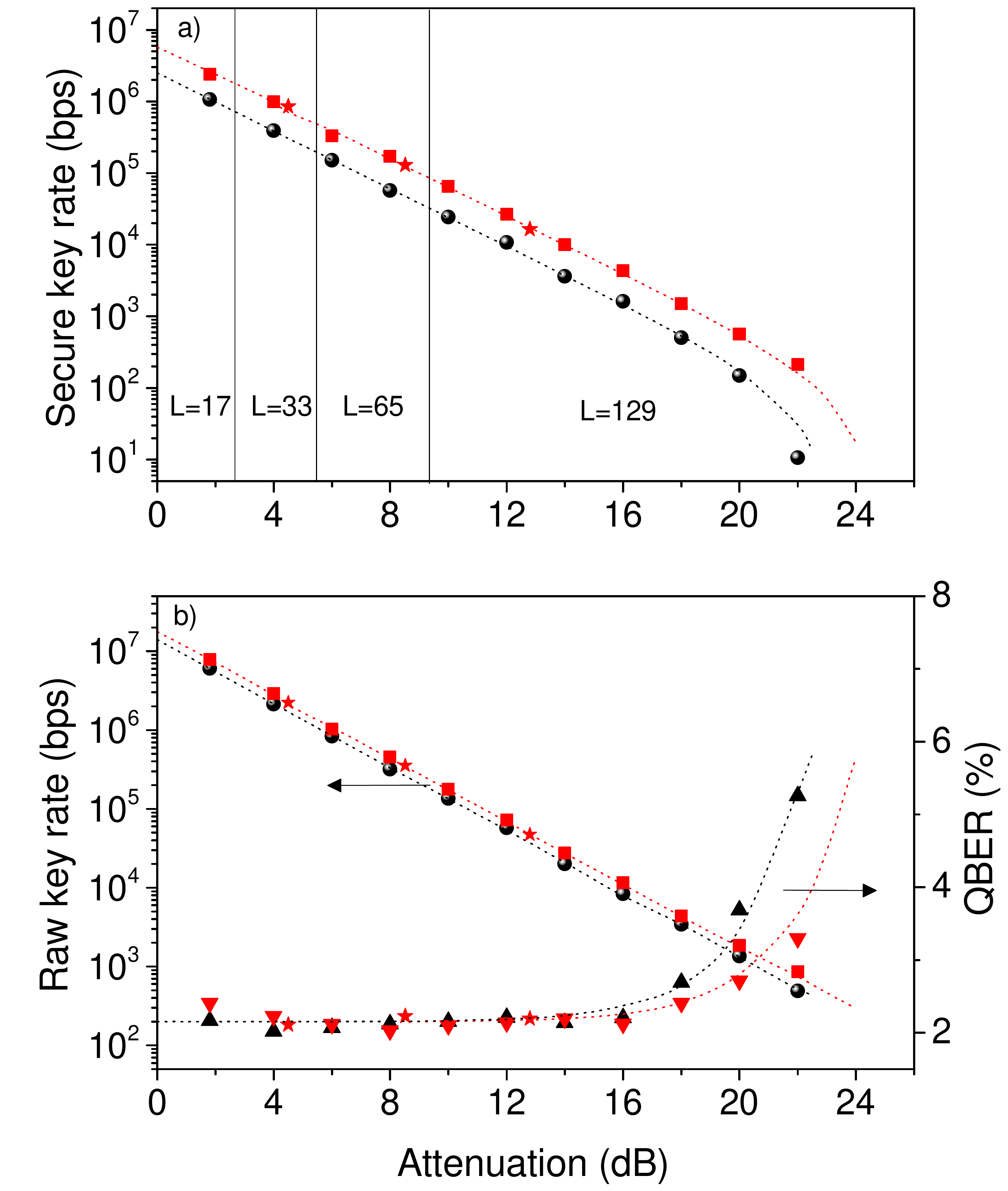}
    \caption{\textbf{Protocol results.} Experimental (symbols) and simulated (dotted lines) key rates and error rates for optical attenuators and real optical fiber (stars) as the quantum channel. \textbf{(a)} The secure key rates are shown for DQPS (above, red squares) and BB84 (below, black circles). The block sizes used at each distance for DQPS are overlaid. \textbf{(b)} The raw key rates for DQPS (above, red squares) and BB84 (below, black circles). The QBERs are also displayed for DQPS (red downwards triangles) and BB84 (black triangles).}
    \label{fig:keyRates}
\end{figure}

We now show the resulting secure key rate dependence on channel attenuation (red symbols), Fig.~\ref{fig:keyRates}a.
Also plotted for comparison are results for the BB84 protocol (black symbols).
We can produce secure keys up to a channel attenuation of 22~dB, which is equivalent to 110~km of standard optical fiber at 1550~nm, using the DQPS protocol.
We also record the data for real fiber lengths of 20, 40 and 60~km, which are well aligned with the simulated results and other experimental data.
The secure key rate for DQPS, which reaches megabit per second rates, is higher than BB84 for all channel attenuations, and the DQPS protocol is able to produce secure keys at longer distances.
The base QBER is low, at  an average of 2.15~\%.
The QBER rises for large attenuations due to the increasing influence of detector dark counts, limiting the secure key rate.

\begin{figure}[ht]
    \centering
    \includegraphics[width=\linewidth]{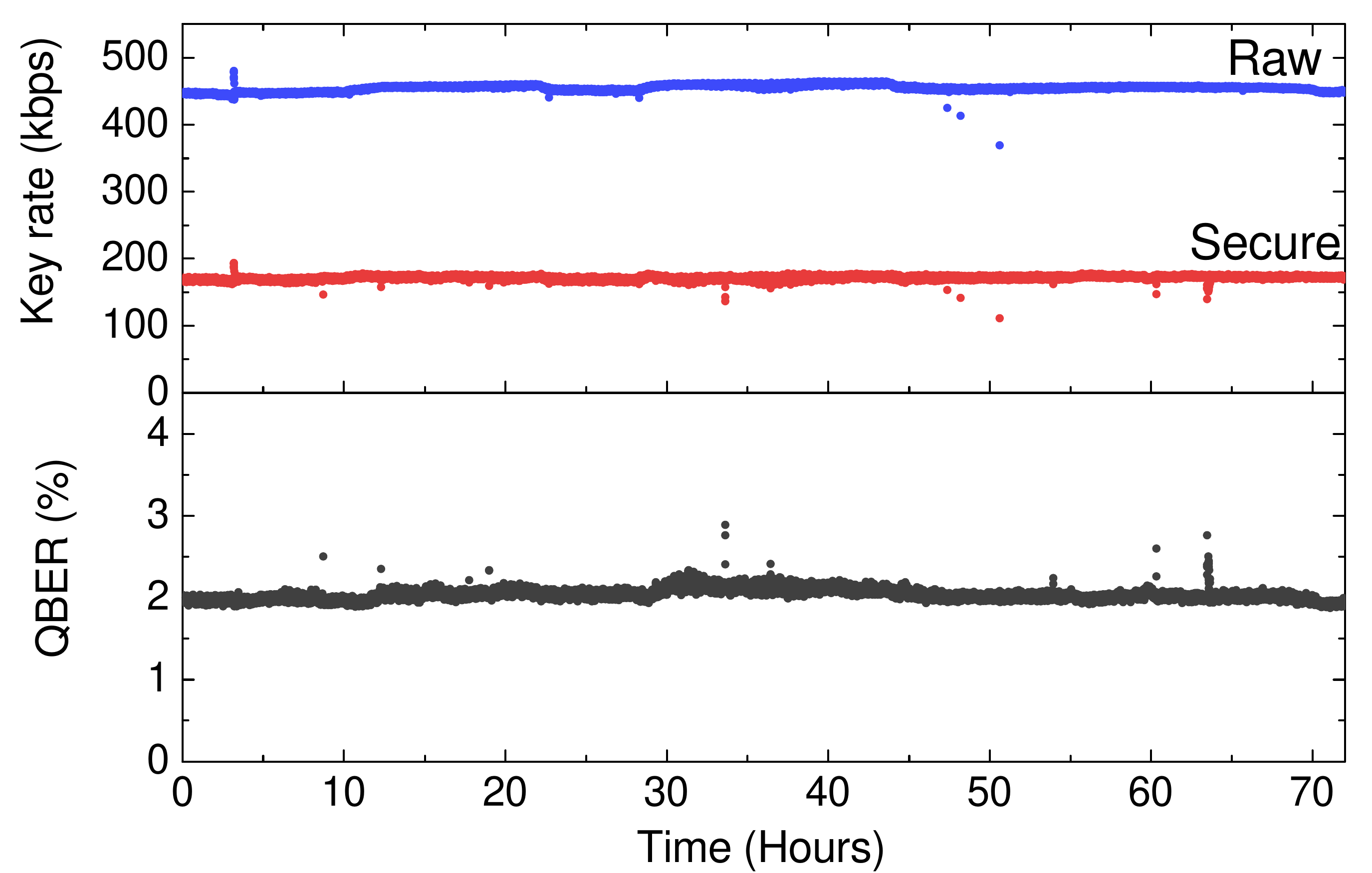}
    \caption{\textbf{Stability.} The extrapolated key rates are shown alongside the QBER for three days of uninterrupted transmission at 8 dB channel attenuation. L=65, the mean photon number is 0.00722 photons per pulse and the integration time is 4 seconds. Counts and QBER are measured in the Y basis, and we assume these are the same in the X basis.}
    \label{fig:stability}
\end{figure}

The stability of the free-running system with no active feedback is shown in Fig.~\ref{fig:stability}.
The average QBER is 2.03~$\pm$~0.06~\%, enabling an average secure key rate of 171.272~$\pm$~2.645~kbps, with no drops in secure key over the entire period of 72~hours continuous operation.
This would amount to a total of 4.95~Gbits of secure key material to be distributed between Alice and Bob.

Phase encoded BB84 is currently a widely used protocol because of its straightforward implementation.
We have shown that the DQPS protocol is able to extend the obtainable BB84 key rates by a factor of 2.71 with no consequences on the experimental complexity.
As with BB84, the DQPS protocol also offers unconditional security.
We note that the performance of the BB84 protocol has been significantly enhanced using decoy states~\cite{Lo_decoy_2005,Wang_beating_2005}, at the expense of implementation simplicity because intensity modulators are required.
However, we believe the decoy-state technique can equally enhance the performance of the DQPS protocol, given that the BB84 protocol is just a special case of the DQPS protocol (L=2).

The promising properties of the transmitter are also highlighted by the experimental results.
The base QBER of 2.15~\% is lower than many other QKD implementations~\cite{Dixon_continuous_2010,Lucamarini_efficient_2013}, and allows us to achieve excellent key rates.
A simple change in input patterns to the master and slave laser allow the transmitter to implement both phase and intensity modulated QKD protocols.
This paves the way to single systems that can choose a protocol based on particular clients, and also easily adapt to new protocols.
Additionally, many current QKD transmitters require time consuming active feedback mechanisms to ensure the system remains stable~\cite{Yuan_continuous_2005}, however the stability data presented in Fig.~\ref{fig:stability} shows that this is unnecessary in the current implementation, giving accurate phase modulation over three days.
The secure key rates of the DQPS protocol over three real-fiber distances also align well with the theoretical values and those obtained using an optical attenuator, proving the system's performance in a realistic scenario.

In summary, we have shown the enhanced performance of the DQPS protocol over the phase-encoded BB84 protocol. 
We have also shown the high stability of the directly phase-modulated transmitter over three days with no need for complicated active feedback mechanisms. 
We conclude that the simplicity and stability of the transmitter, alongside the high secure key rate provided, show that the DQPS protocol is promising for real-world applications in the near-future. 

\section*{Supplementary Material}
See supplementary material for the autocorrelation results of a DQPS pattern over 100 lags.

\section*{Acknowledgements}
G.L.R gratefully acknowledges financial support from the EPSRC CDT in Integrated Photonic and Electronic Systems and Toshiba Research Europe Limited.

\bibliographystyle{apsrev4-1}
\bibliography{FullBibDQPS}

\end{document}